# A Fast MR Fingerprinting Simulator for Direct Error Estimation and Sequence Optimization

Siyuan Hu, Stephen Jordan, Rasim Boyacioglu, Ignacio Rozada, Matthias Troyer, Mark Griswold, Debra McGivney, Dan Ma

*Abstract*—MR Fingerprinting is a novel quantitative MR technique that could simultaneously provide multiple tissue property maps. When optimizing MRF scans, modeling undersampling errors and field imperfections in cost functions will make the optimization results more practical and robust. However, this process is computationally expensive and impractical for sequence optimization algorithms when MRF signal evolutions need to be generated for each optimization iteration. Here, we introduce a fast MRF simulator to simulate aliased images from actual scan scenarios including undersampling and system imperfections, which substantially reduces computational time and allows for direct error estimation and efficient sequence optimization. By constraining the total number of tissues present in a brain phantom, MRF signals from highly undersampled scans can be simulated as the product of the spatial response functions based on sampling patterns and sequence-dependent temporal functions. During optimization, the spatial response function is independent of sequence design and does not need to be recalculated. We evaluate the performance and computational speed of the proposed approach by simulations and in vivo experiments. We also demonstrate the power of applying the simulator in MRF sequence optimization. The simulation results from the proposed method closely approximate the signals and MRF maps from in vivo scans, with 158 times shorter processing time than the conventional simulation method using Non-uniform Fourier transform. Incorporating the proposed simulator in the MRF optimization framework makes direct estimation of undersampling errors during the optimization process feasible, and provide optimized MRF sequences that are robust against undersampling factors and system inhomogeneity.

## I. INTRODUCTION

Magnetic Resonance Fingerprinting (MRF) [1] is a quantitative MR technique that enables simultaneous mapping of multiple tissue properties from a single scan within a clinically feasible time. An MRF scan typically consists of hundreds to thousands of time points where radio-frequency pulses are applied to tip the proton spins in tissues. Each pulse employs varying flip angles and is separated with varying repetition times (TR) to generate distinct signal evolutions for different tissues. A dictionary is generated following Bloch equations to foresee the signal behavior of all possible combinations of tissue properties under the impact of the applied pulse sequence. The quantitative tissue property maps are obtained by matching the acquired signal at each pixel/voxel against all the dictionary entries. To acquire sufficient information for accurate and robust tissue parameter estimation, more images are always preferred leading to long scans. Various techniques have been adopted in MRF to speed up the acquisition time, such as k-space undersampling with variable-density trajectories [2]–[4]. The accelerated MRF scans with a 3D whole-brain coverage can be achieved within 5 mins with 1x1x3 mm or 1 mm isotropic resolutions in recent studies [3], [4]. However, pushing the MRF scans to submillimeter resolutions with a large volumetric coverage will require much longer scan times, leading to patient discomfort, higher expenses, and greater chance of motion artifacts. Although there exists an overall tradeoff between short scan duration and high measurement accuracy, pulse sequence optimization could potentially enable short MRF scans to achieve equivalent quality of tissue property mapping as that of long MRF scans.

There have been previous studies to optimize the sequence designs of MRF for shorter scan times by minimizing criterion-specific cost functions [5]–[7]. For example, Zhao et al. [5] used Cramér-Rao Bound principles to minimize variance of parameter estimations due to random noise. Cohen et al. [6] accounted for orthogonality between dictionary entries. Sommer et al. [7] looked at correlations between noise contaminated signals and global dictionary signals. In many of these studies, Gaussian random noise is used to represent both measurement noise and undersampling artifacts. Although such an approximation works in cases of random sampling with many MRF time points, it may not be accurate in representing aliasing artifacts from undersampled trajectories, such as

The authors would like to acknowledge funding from Siemens Healthineers, Microsoft and NIH grant EB026764-01 and NS 109439-01 (*Corresponding author: Dan Ma*)

Siyuan Hu, Debra McGivney and Dan Ma are with the Department of Biomedical Engineering, Case Western Reserve University, Cleveland, OH 44106 USA (e-mail: siyuan.hu@case.edu; debra.mcgivney@case.edu; dan.ma@case.edu).
Stephen Jordan and Matthias Troyer are with Microsoft Corporation, Redmond, WA 98052 USA (e-mail: stephen.jordan@microsoft.com; mtroyer@microsoft.com).
Rasim Boyacioglu and Mark Griswold are with the Department of Radiology, Case Western Reserve University, Cleveland, OH 44106 USA (e-mail: rasim.boyacioglu@case.edu; mark.griswold@case.edu).
Ignacio Rozada is with 1QBit Information Technologies Inc., Vancouver, BC V6E 4B1 (e-mail: ignacio.rozada@1qbit.com).

commonly used spiral and radial trajectories [8] from short sequences. In addition, the signal is typically modeled in an ideal case and does not take system imperfections into account, which further simplifies the conditions as compared to an actual acquisition. As a result, the optimized MRF sequences usually suffer from limited robustness and performance gain when applied to actual in vivo scans that typically employ high acceleration factors.

To improve the robustness and performance of the optimization, artifacts from actual scans have to be considered. However, simulation of undersampling artifacts is rarely introduced in optimization routines, mainly due to its high computational cost. Unlike random errors, the aliasing artifacts are spatially dependent, so the artifacts are best approximated by first-principles simulation using explicit phantom maps of realistic tissue compositions and spatial distributions, which are undersampled using a chosen acquisition trajectory, and then reconstructed to obtain aliased maps of the phantom. Directly minimizing the errors of the resulting quantitative maps of tissue properties is a more predictive approach, as compared to the implicative cost functions introduced in previous works that measure signal magnitude and/or orthogonality. However, such error estimation requires two necessary steps. One is image reconstruction that involves non-uniform fast Fourier transform (NUFFT) [9] with additional steps of gridding and density compensation; the second is quantitative mapping that requires dictionary simulation and pattern matching. These steps take the majority of the computing time. Although several reconstruction toolboxes have been developed for efficient image reconstruction from arbitrary sampling trajectories, large computational power is still needed for the purpose of sequence optimization. Specifically, the cost functions of many sequence candidates have to be evaluated to approach the optimal solution, and the artifacts from hundreds of images need to be computed in each sequence candidate. In our example of MRF optimization, around 44,000 iterations are required to find the optimal MRF sequence using simulated annealing (SA) algorithm [10] (the maximum iterations in the SA algorithm are typically greater than 100, more complex objective functions require larger amounts of iterations of searching). Repeating the NUFFT and gridding calculations at each iteration in such an optimization problem is therefore computationally expensive and impractical.

Here we propose a fast MRF simulator to address the computational challenge of repeatedly simulating aliased images during MRF sequence optimization. The proposed simulator provides simulated signals that best approximate the signals from actual in vivo scans by considering undersampling factors and system imperfections. The concept is similar to the partially separable models used in image reconstruction, where the time-resolved reconstructed images are represented as a product of low rank spatial and temporal basis functions [11]–[14]. Here, we use this concept to simulate aliased images of actual scans. By constraining the total number of tissues considered in the optimization, the spatial response from non-Cartesian undersampling can be formulated and precalculated separately from the temporal signal variation during simulations. As a result, a set of aliased images can be simulated as a direct product of the precalculated spatial response functions and time-varying signal intensities. This is especially time-saving for scan optimization when the cost functions from tens of thousands of sequence patterns are to be evaluated, because the spatial response functions that include NUFFT computation do not need to be re-calculated for each evaluation.

We show that the proposed method is capable of accelerating simulations while approximating the aliasing artifacts from in vivo scans. In this study, we demonstrate using the simulator to model the aliased brain images, but the method can be applied to any anatomical structures with different tissue distributions. We further show an example of incorporating the proposed simulator into the MRF sequence optimization framework, which helps to improve the robustness of optimized MRF scans against undersampling and system imperfections. In general, any MRI applications that require simulation of a large number of signals from different scan scenarios including varying sampling trajectories, excitation patterns and system imperfections, such as scan optimization and deep-learning training, could benefit from the proposed approach.

## II. Theory

The signal from a pixel in an MRI scan typically results from a distribution of different tissues. However, in the context of an optimization, we can approximate the anatomy as being made up of a finite number of tissues of interest. In this case, assuming there is no motion during the scan, the spatial basis of the pixel remains static and separable from time-varying signal intensities and we can define the signal from a pixel, $s(t)$ as:

$$s(t) = \sum_{i=1}^{J} p_i d_i(t) \quad (1)$$

Where $J$ is the total number of possible tissue types in the pixel, $p_i$ is the volume fraction of a tissue type $i$ in the pixel, and $d_i(t)$ indicates the signal evolution of the tissue type. Assuming that all tissues of the same type are uniform, and yield identical signal patterns, (1) can be generalized to express the signals of fully sampled images:

$$I_{full}(x,y,t) = \sum_{i=1}^{J} P_i(x,y) d_i(t) \quad (2)$$

where $P_i(x,y)$ is the volume fraction of tissue type $i$ in each pixel. In this study, volume fractions of each tissue in the masks have a value of either 1 or 0, but partial volume effects represented by volume fraction distributions between 0 and 1 can also be easily simulated. Based on this representation, the undersampled image can be written as:

$$\begin{aligned} I_{us}(x,y,t) &= F_{us}^{-1} K F_{full}(I_{full}) \\ &= \sum_{i=1}^{J} F_{us}^{-1} K F_{full} P_i(x,y) d_i(t) \end{aligned} \quad (3)$$

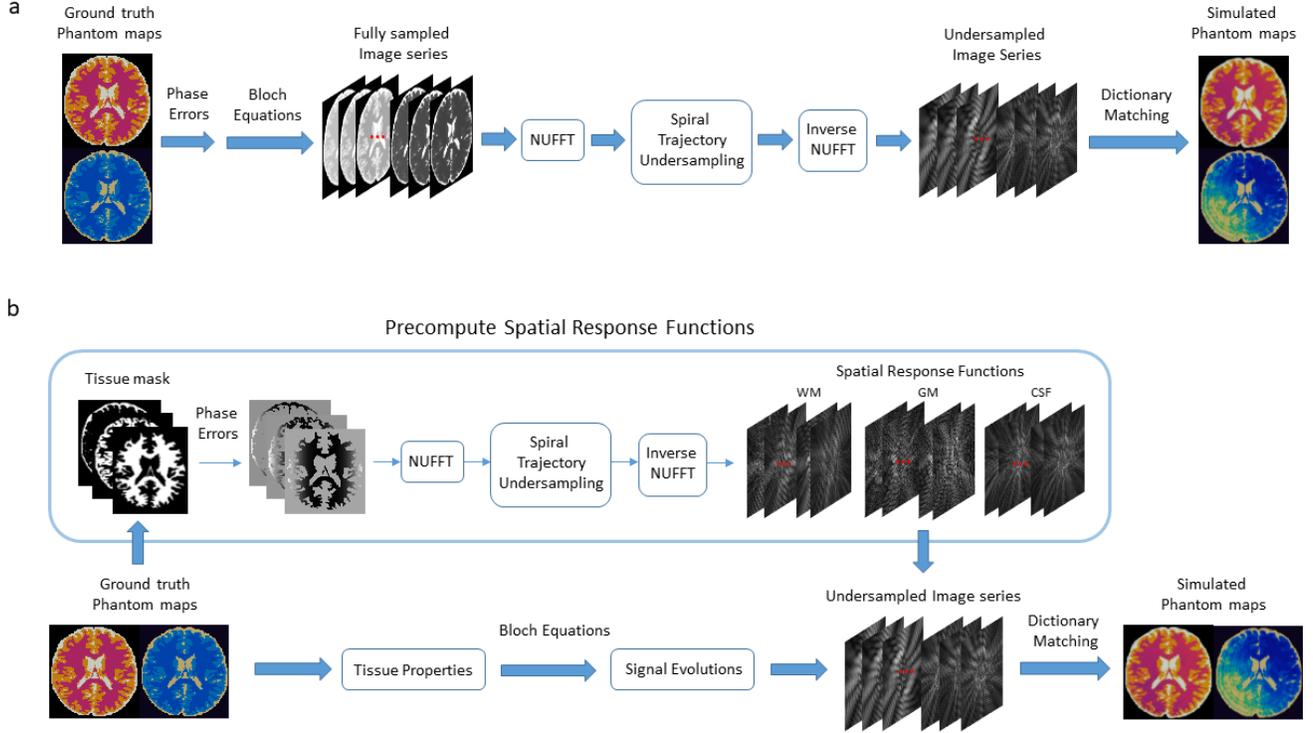

Fig. 1. The workflow structure of (a) conventional first-principle simulation method and (b) the proposed fast image series simulator. The steps to simulate undersampled image series from the ground truth reference maps are connected in series in the conventional method, but the proposed simulator divides these calculations into two parallel processes. One is the spatial response functions that contain all static spatial information and NUFFT calculations to be precomputed. The other derives signal evolutions in temporal domain that varies in different scans with different acquisition parameters. Combining the spatial response function and signal evolutions gives undersampled image series. In real applications, only the signal evolutions need to be re-evaluated for simulating images acquired using different sequences. Quantitative mapping to obtain T1 and T2 maps from the image series are identical in both methods.

where $F_{us}$ and $F_{full}$ are undersampled and fully sampled NUFFT operators respectively, and $K$ is the undersampling pattern in k-space. Since both the NUFFT and undersampling take place in the spatial frequency domain and are independent from the signal evolution $d_i(t)$ which is governed by spin dynamics, (3) can be written as:

$$I_{us}(x,y,t) = \sum_{i=1}^{J} \Psi_i(x,y) d_i(t)$$

with

$$\Psi_i(x,y) = F_{us}^{-1} K F_{full} P_i(x,y) \quad (4)$$

Assuming $K$ is determined a priori, $\Psi_i$ can be precalculated using tissue distributions and for each k-space trajectory. Incorporating more spatial-dependent factors in the model can be achieved by adding extra factors in the $\Psi_i$ matrix. For example, to model the effect of static background phase $\theta_i(x,y)$ on top of undersampling, one could simply modify $\Psi_i$ to be:

$$\Psi_i(x,y) = F_{us}^{-1} K F_{full} P_i(x,y) e^{j\theta_i(x,y)}$$

Given the known $\Psi_i$, the only calculated component for the simulation of undersampled images are the time-varying magnetization intensities, which follow spin evolutions manipulated by sequence parameters.

### III. PROPOSED FRAMEWORK

The computations involved in the proposed framework of fast MRF simulator could be divided into three major steps: spatial response functions, temporal-varying signals and dictionary, and quantitative mapping. Fig. 1 summarizes the overall workflow of this framework (Fig. 1b) in comparison with the conventional first-principle simulation method (Fig. 1a). The following sections discuss these steps in details.

Assuming signals from MR brain scans are governed by a few major tissue components, we created a numeric brain phantom with matrix size of 256x256 based on an MNI brain template [15]. Specifically, for the ease of MRF sequence optimization, the brain phantom is simplified and segmented with 3 tissue types: white matter (WM), gray matter (GM) and cerebrospinal fluid (CSF), as shown in Fig. 2a. We propose to simulate aliased images and estimate MRF maps using the steps described in the subsequent sections.

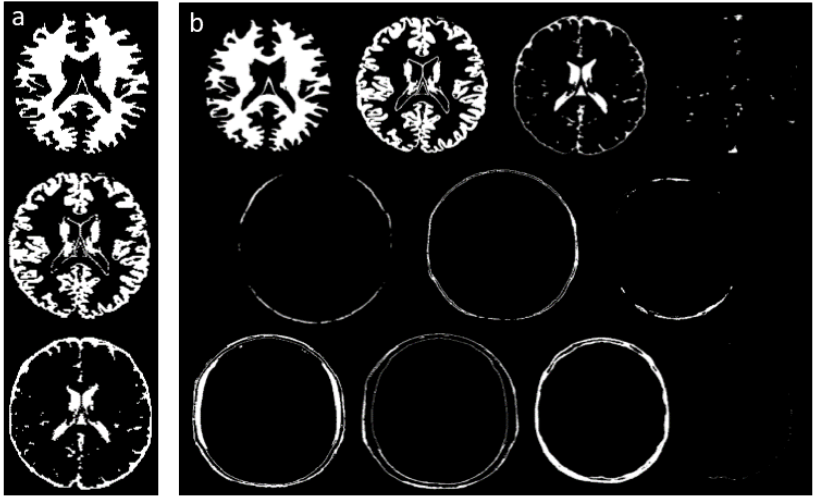

Fig. 2. Segmented digital brain phantoms. (a) simplified 3-tissue brain phantom, up to down are: white matter (WM), gray matter (GM) and cerebrospinal fluid (CSF). (b) 11-tissue brain phantom that consists of (left to right, up to down): WM, GM, CSF, blood vessel, fat, tissues around fat, bone marrow, muscle, skin around muscle, skull, dura matter

### A. Precompute spatial response functions

In our actual MRF scans, images at each time point are undersampled using one spiral arm out of 48 spiral trajectories, as described in Ma et al [1]. In the simulation, we first fully sampled the masks of each tissue segment in k-space using 48 variable density spiral trajectories. All the spirals were needed to fully sample the k-space to obtain $F_{full}$. Retrospective undersampling was performed by sampling each tissue mask with one of the spirals for acceleration, which gives $F_{us}$. Density compensation and inverse NUFFT were then applied to the undersampled k-space corresponding to each tissue mask, resulting in 48 highly undersampled masks for each tissue. These undersampled images of tissue segment masks in spatial domain were referred as "spatial response functions" for each tissue, denoted $\Psi_{wm}(x,y,S)$, $\Psi_{gm}(x,y,S)$, $\Psi_{csf}(x,y,S)$, as functions of position $(x,y)$ and spiral interleaf $S$, $S=1...48$. The spatial response functions described the change of spatial distribution of each tissue segment appearing on a undersampled image, in response to the undersampling operations performed with a particular acquisition trajectory. The same tissue mask would yield different spatially aliased distributions when it is undersampled by different spiral interleaves. During an actual acquisition, the 48 spiral interleaves were linearly ordered and repeated for every 48 images, so we needed 48 spatial response functions for each tissue type.

In this study, we also added a background B0 phase in the simulation to account for field variations and inhomogeneity. Background phase maps were synthesized with a parabolic pattern ranging from -π to 2π. The phase map was multiplied by the brain phantom mask in the spatial domain prior to performing undersampling calculations. As the phase variations could possibly occur in any direction in an actual MR scan, we show phase maps in 4 different directions in the simulations (Fig. 5).

### B. Generate signal evolutions and dictionary

In this work, we simulated the MRF scan acquired using an example FISP-MRF sequence described in Jiang et al [16] with 480 time points. We implemented Bloch equations based on the FISP-MRF sequence to generate a dictionary that predicts the signal evolutions of a set of combinations of T1 and T2. The dictionary entries covered a range of 2-3000 ms for T1 and 2-2000 ms for T2. The signal evolutions for the three tissue types, $d_{wm}(t)$, $d_{gm}(t)$, and $d_{csf}(t)$, were derived similarly following Bloch equation. Specifically, each tissue was registered with relaxation times typical in a healthy human [17]: WM (T1 = 800 ms, T2 = 40 ms), GM (T1 = 1400 ms, T2 = 60 ms) and CSF (T1 = 3000 ms, T2 = 500 ms).

### C. Simulate reconstructed images and MRF maps

Given the spatial response functions $\Psi_i$ and signal evolutions $d_i(t)$ calculated for WM, GM, and CSF in section A and B, the final reconstructed images $I_{us}(x,y,t)$ from the undersampled data could be obtained from Eq. 4. For each tissue segment, we computed the direct product of the linearly ordered spatial response functions and time-varying signal evolutions. Note that we had signal evolutions of 480 time points, and 48 spatial response functions derived from 48 spiral interleaved that are repeatedly used during the MRF scan, so these spatial response functions were sequentially replicated to match the length of signal evolutions before the multiplication. It leads to 3 spatiotemporal terms that correlates to the three tissue types. Linear combination of these terms yielded a series of aliased images as a function of time, which were then matched to the dictionary to generate quantitative T1 and T2 maps. The simulated maps derived here could be compared with ground truth to estimate the measurement errors, which could be used as the cost function in the context of sequence optimization.

### D. MRF sequence optimization

To demonstrate the power and practicality of the proposed method, we applied this simulator in MRF sequence optimization. We constructed a cost function that calculates the errors of the quantitative maps using the simplified 3-tissue brain phantom to optimize MRF scans. The optimization seeks the MRF sequence pattern that yields minimal T1 and T2 errors of GM, WM and CSF in the simulated aliased MRF maps.

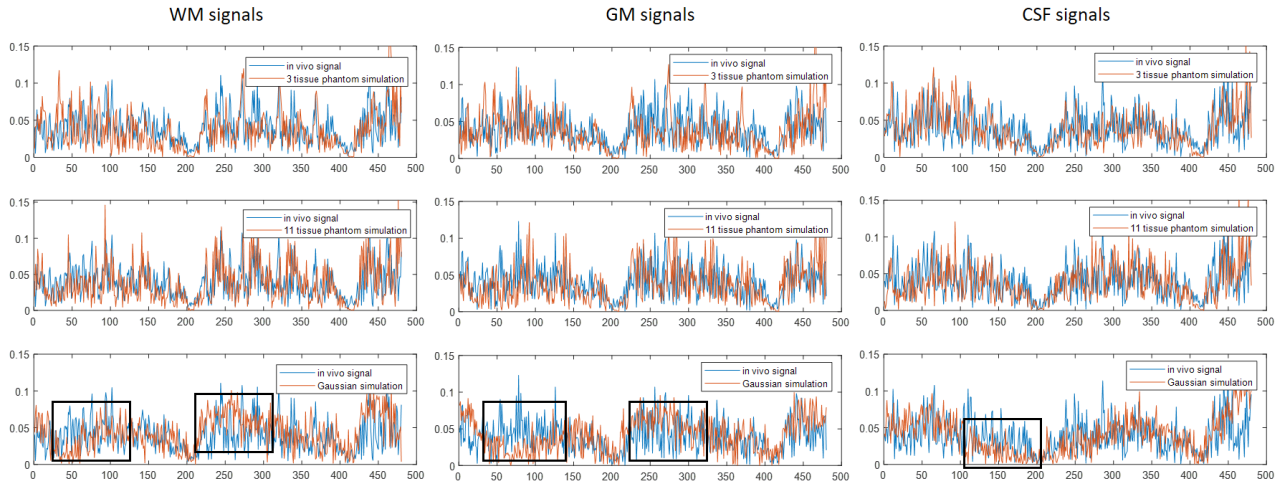

Fig. 3. Normalized MRF signals simulated using the proposed method with the 3-tissue phantom (row 1) and the 11-tissue phantom (row 2), Gaussian model (row 3) as compared to in vivo signals. Shown in the red are the simulated signal evolutions for each of WM, GM and CSF. The blue signals were extracted from a pixel within the corresponding tissue region that have the same T1 and T2 values in an in vivo scan. The scale and shape of the aliasing artifacts from both 3-tissue phantom based model and 11-tissue phantom based model closely matches the measured signal

Specifically, we simulated T1 and T2 maps acquired by a candidate sequence under undersampling and background phase using the proposed simulation procedures. Note that the simulation performed here was obtained with phase variation in only one direction, instead of all 4 directions. The root-mean-square T1 and T2 errors of WM, GM, and CSF tissue segments on simulated maps were estimated in the cost function. The cost function also evaluated the robustness of the MRF sequence against random noise as described in Kara et al. [18], which derived quality factors to indicate the likelihood of measurement errors of T1 and T2 for each tissue type due to thermal noise. In our proposed optimization framework, the cost function was a function of the simulated errors and quality factors of all the three major tissue types, which was then divided by a scan-time penalty.

While our framework could optimize MRF scans of any length, here we demonstrate optimizing MRF sequences of 480 time points. To handle such non-convex and high-dimensional problem, the optimization was carried out by stochastic Monte Carlo [19] and simulated annealing [10] algorithms. Further details of optimization implementation are described in [20].

## IV. METHODS

### A. Validation

To assess the validity of the proposed brain phantom to approximate the signals acquired in vivo, we also created an 11-tissue brain phantom (Fig. 2b). This phantom was used to simulate the image series and MRF maps following the same procedure as described in section III.A-C. In addition to WM, GM, and CSF, it contains segments of blood vessels (T1 = 1600 ms, T2 = 100 ms), fat (T1 = 360 ms, T2 = 70 ms), tissues around fat (T1 = 500 ms, T2 = 70 ms), bone marrow (T1 = 500 ms, T2 = 70 ms), muscle (T1 = 800 ms, T2 = 48 ms), skin around muscle (T1 = 560 ms, T2 = 320 ms), skull (T1 = 0 ms, T2 = 0 ms) and dura mater (T1 = 0 ms, T2 = 0 ms) [15]. To make this 11-tissue phantom more realistic, we also integrated partial volume effects. The volume fraction mask $P_i(x,y)$ for each tissue type was no longer binary, but has values ranging from 0 to 1. To implement the fast simulator using this complex phantom, 8 more spatial response functions and signal evolutions were calculated for these additional tissue segments and tissue types. Simulation results from both phantoms were

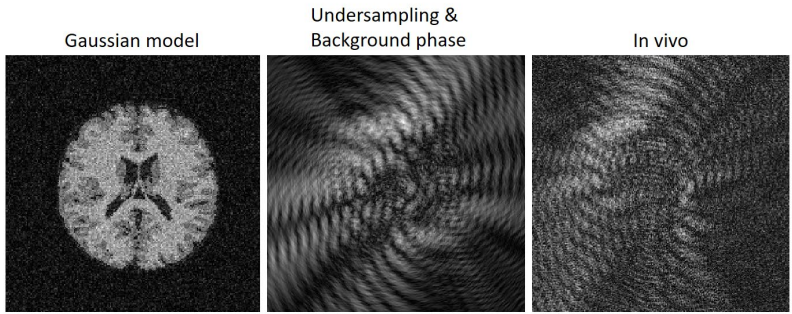

Fig. 4. Simulations of the aliased images using the Gaussian model and the proposed method that incorporated undersampling and background phase, as compared against the image acquired in an actual in vivo MRF scan. The simulation from the proposed method could replicate the aliasing features in the undersampled in vivo image.

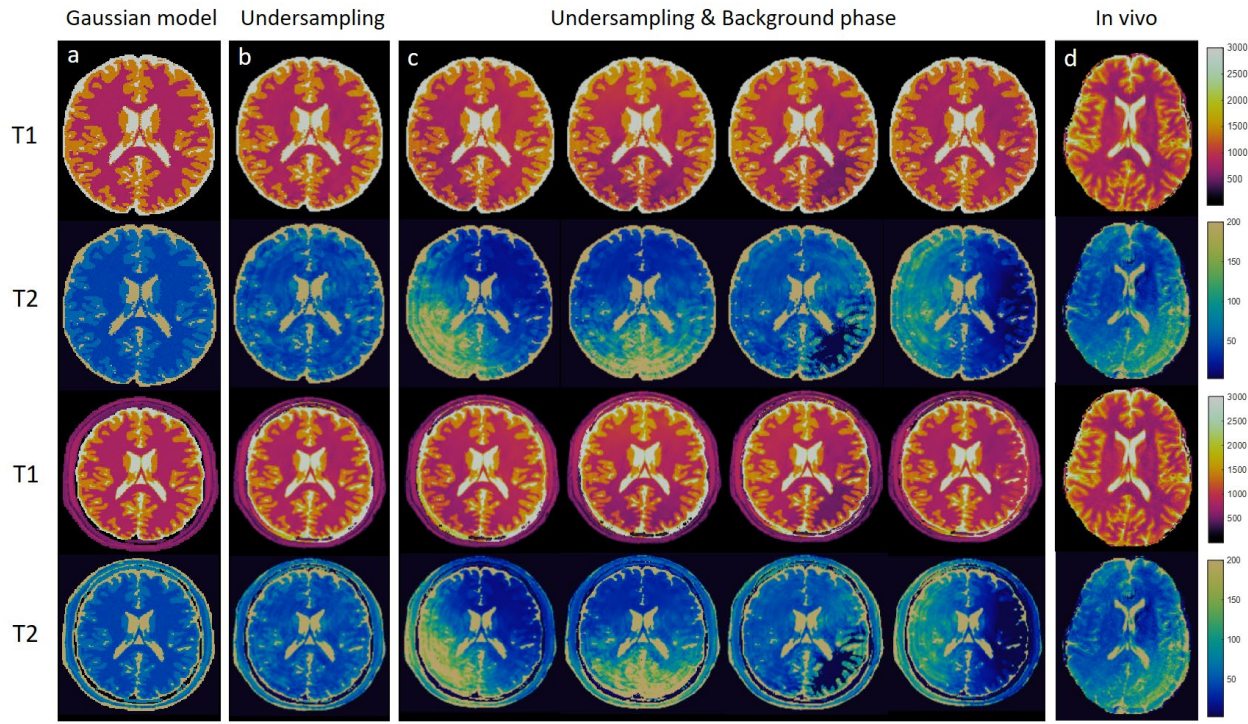

Fig. 5. T1 and T2 MRF maps simulated using the 3-tissue phantom (row 1,2) and the 11-tissue phantom (row 3,4) that incorporates (a) Gaussian random noise, (b) single shot spiral undersampling, (c) undersampling and phase variations in 4 different directions, and (d) in vivo MRF scan acquired by the same sequence. Only the simulations in (c) best approximate the shading artifacts observed in in vivo scans. The simulations generated from the simplified 3-tissue phantom exhibit the same shading features in the images

compared against in vivo signal evolutions and in vivo MRF maps of human brain.

Furthermore, we modeled the undersampling errors using Gaussian noise as a comparison. Gaussian white noise with SNR = 9dB were added to the clean signals of WM, GM and CSF from the dictionary described in Section III.B. Note that to meet the assumption that aliasing errors are dependent upon MRF signal magnitude [7], [8], here the signal power were used to determine the Gaussian noise level based on the given SNR value. The Gaussian noise contaminated signals were then matched to the dictionary to generate T1 and T2 maps. The results were compared to the simulations derived from the proposed method using both the 3-tissue phantom and the 11-tissue phantom. The signal evolutions and MRF maps simulated from the Gaussian model were also compared with those acquired in in vivo scans.

We compared the computational speed between the proposed method and the conventional approach using NUFFT reconstruction for each image. The conventional method (Fig. 1a) was also implemented using the same brain phantom, spiral trajectories, MRF sequence, tissue property values and dictionary. Specifically, a series of brain phantom images acquired from 480 TRs were first simulated by applying the Bloch equations. Each of the 480 images was then transformed into k-space via NUFFT and undersampled with the single shot spiral trajectories of the same linear ordering. After reconstruction of the undersampled k-space data using inverse NUFFT, dictionary matching was performed to obtain MRF maps. Both the proposed and the conventional methods were implemented using MATLAB version 2019b.

### B. *In vivo experiment*

As a reference to the simulations in section IV.A, an in vivo brain scan from a healthy volunteer was performed in a Siemens 3T Skyra scanner using the example FISP-MRF sequence with 480 time points. The optimized MRF sequences of 480 TRs were also validated with in vivo scans on a Siemens 3T Vida scanner. In vivo scans were conducted with the approval from the Institutional Review Board. All the scans were acquired with a FOV of 300x300mm$^2$, matrix size of 256x256, using single shot spiral trajectories with a undersampling factor of 48.

Image reconstructions for all in vivo scans were performed using non-uniform fast Fourier Transform and direct dictionary matching.

## V. RESULTS

### A. *Simulation*

Fig. 3 compares the signal evolutions simulated from the 3-tissue brain phantom, the 11-tissue brain phantom and Gaussian noise model with signals acquired in an actual in vivo MRF scan. Signal evolutions of each of the three representative tissue types, GM, WM and CSF, from the same brain region were compared. All the signals shown in the figure were normalized reveal brain structures. In Fig. 5a, the MRF maps were derived to illustrate the relative magnitude of aliasing errors in the context of signal intensities. The signals from the in vivo scan were severely corrupted by aliasing artifacts. The magnitude of the aliasing errors was notably high compared to the signal level, leading to low SNR and errors in dictionary matching. In the simulations obtained from both the 3-tissue phantom (row 1) and the 11-tissue phantom (row 2), the scale of undersampling artifacts represented in the proposed method closely matches that from the in vivo scans for all tissue types. Moreover, the general shapes of in vivo signals were often slightly shifted from the corresponding predictions from the dictionary due to system imperfections during data acquisition. These simulated signals, even the ones generated using the simple segmented 3-tissue phantom, were able to replicate the aliasing patterns and the overall trends of these biased from the in vivo signals. This validates our model for close approximation of in vivo conditions. By using the Gaussian model-based simulation (row 3), one can also produce errors at the same magnitude level as real undersampling errors by tuning the SNR value. However, the Gaussian model fails to capture the error patterns and general shape of in vivo signals, as shown in the highlighted regions in Fig. 3. Since Gaussian white noise is an unbiased noise model, it is not able to reproduce the aliasing features of in vivo signals that were acquired under a non-ideal measurement environment.

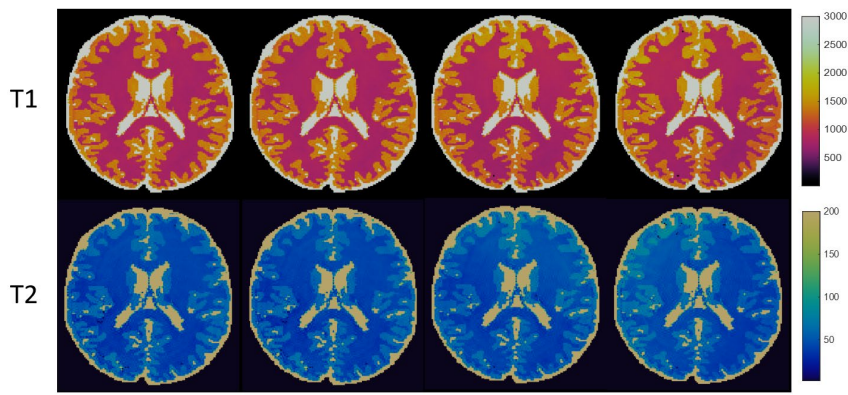

Fig. 6. Simulated MRF maps of an optimized sequence that incorporates undersampling and phase variations in 4 different directions. Although the sequence was optimized towards building its robustness against phase variation in only one direction, the result optimized sequence also demonstrates comparable performance under phase variations in other directions.

Fig. 4 demonstrates an undersampled and aliased images acquired in an in vivo MRF scan, and its comparison with the simulated images from both the Gaussian simulation model and the proposed method. The actual undersampled image is subject to severe aliasing artifacts that exhibit highly spatially dependent "ringing" patterns and corrupts the brain structures in the image. These image features can also be observed in the simulation generated from the proposed approach. In contrast, the Gaussian model is not able to represent such spatial dependence in the error patterns.

Fig. 5 shows the MRF maps computed from the aliased images that were simulated with different approaches using the 3-tissue phantom and the 11-tissue phantom, as compared to the actual in vivo MRF scans. Even though the signals are highly aliased, the resulting MRF maps in Fig. 5 could still clearly from the signal evolutions simulated based on the Gaussian noise model, and Fig. 5b shows the MRF maps of the model only representing undersampling factors. As comparison, both undersampling and background phase were incorporated in the simulation in Fig. 5c. Fig. 5d shows a MRF scan acquired in vivo as a reference. Since the in vivo scan was performed with a shortened sequence of 480 time points, quantitative mapping using the short, highly undersampled, and biased signal evolutions are likely to cause mismatches to wrong dictionary entries. As a result, the resulting MRF maps of the in vivo scan, especially the T2 map, are subject to severe ringing artifacts and shading artifacts, where a large region on the image is mostly overestimated and the opposite side is underestimated.

In the Gaussian-model based simulation (Fig. 5a), both T1 and T2 maps that were reconstructed from the Gaussian-noise contaminated images failed to mimic these artifacts. In the undersampling-only simulations shown in Fig. 5b, the T1 maps of the brain phantom were minimally affected by undersampling; however, the T2 maps exhibited ringing artifacts, which were commonly seen in highly undersampled images. The T2 map from the 3-tissue phantom based simulation suffered more from ringing artifacts than the T2 map from the 11-tissue phantom, since the latter phantom modelled partial volumes at tissue boundaries and contained less sharp edges. When the background phase was added in the simulations together with undersampling (Fig. 5c), the T1 map had slight shading artifacts, while the T2 map showed severe shading and ringing artifacts, which closely matched the artifacts seen in the actual in vivo scans (Fig. 5d) that could not be modeled by Gaussian simulation. While the in vivo MRF maps shown here were subject to shading in the right temporal lobe and occipital lobe, the shading artifacts could appear at any direction in different experiments or when using different scanners. Fig. 5c demonstrates the simulations that were separately computed using 4 background phase error maps in 4 different directions. Note that the same image features are observed in the simulations using both simple 3-tissue phantom and the more realistic 11-tissue phantom. It validates our simple 3-tissue segmented phantom model as equally effective as complex phantoms to represent the real in vivo conditions to approximate the artifacts in actual scans. The 3-tissue phantom thus could be reliable for applications in solving MRF problems.

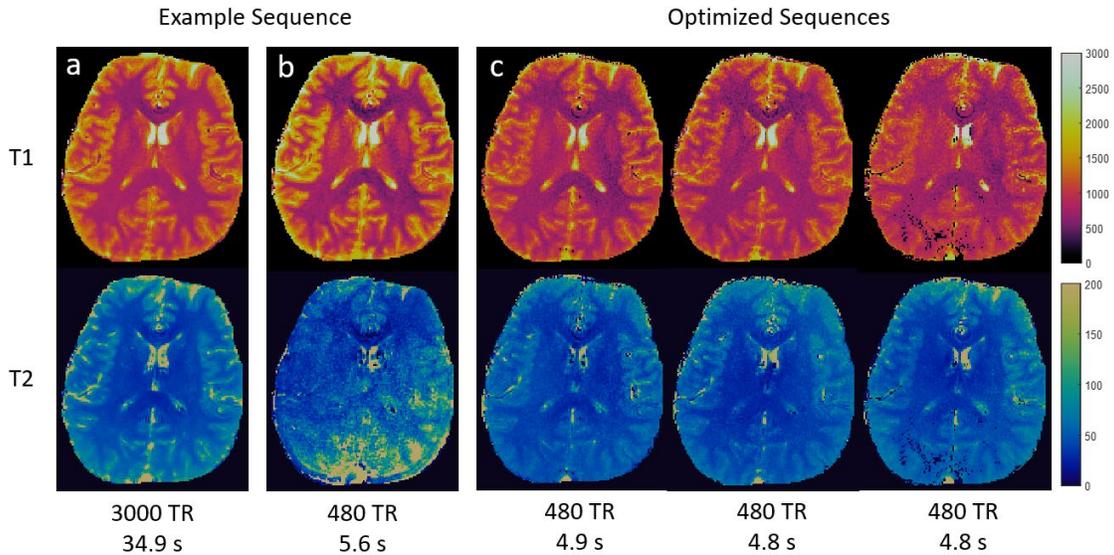

Fig. 7. In vivo T1 and T2 maps acquired from (a) the standard example human-designed sequence of 3000 TRs, (b) the same example sequence truncated to 480 TRs, and (c) multiple optimized scans of 480 TRs. The optimized sequences in (c) are robust against undersampling and system variations that corrupts the shortened human-designed scan.

Calculations using the proposed fast simulator approach were significantly faster than the conventional simulation method. The simulation results of both approaches are exactly equivalent, which can be proven by our mathematical model. Simulation of 480 images with retrospective undersampling from an MRF scan using conventional gridding and NUFFT took 174 seconds, whereas the fast series image series simulator took 1.1 seconds and 2.8 seconds to simulate the same amount of images using the 3-tissue phantom and the 11-tissue phantom, achieving 158-fold and 62-fold accelerations respectively. The computational time of the fast simulator accounts for: 1) generating signal evolutions for each tissue type, and 2) simple matrix multiplications between the spatial response functions and the signal evolutions to compute the final image series. Note that the 11-tissue phantom based framework requires simulating signal evolutions for 8 more tissue types, leading to a longer processing time than the 3-tissue phantom based simulator.

### B. MRF sequence optimization

By incorporating the fast image series simulator into the cost function, we were able to generate optimized MRF sequences that achieve better image qualities within short scan times. In our optimization implementation, the cost function minimizes the shading caused by the phase error in only one direction, instead of evaluating the errors using all 4 phase maps, during the entire optimization. To ensure the optimization was not dependent upon direction of B0 phase, we tested the performance of one optimized sequence under different phase variations on the brain phantom, where all 4 phase maps were separately added in the simulations (Fig. 6). This sequence was optimized based on the phase errors in the third column, however it still shows good resistance against shading in other orientations as well. We can conclude that the robustness of a MRF sequence against phase variations in all directions is built in mutual.

As shown in Fig. 7, we obtained in vivo MRF maps using multiple optimized sequences of 480 time points acquired in different scans (Fig. 7c). Fig. 7a is the gold standard MRF scan acquired using the same human-designed FISP-MRF sequence adopted in previous simulations, but with the full length of 3000 time points. As a reference, we truncated the human-designed sequence to the same length of 480 time points as the optimized ones, leading to a scan time of 5.6 seconds (Fig. 7b). The optimized scans were acquired with scan times around 4.8 seconds. While the truncated human-designed sequence has severe shading artifacts in the occipital lobe, the optimized scans with shorter scan duration do not exhibit these shading artifacts. These reproducible results validate the robustness of the optimized sequences against undersampling effects and phase variations, and the performance consistency of the optimizer in finding reliable MRF sequences.

## VI. DISCUSSION

When evaluating sequence performance for the purpose of scan optimization, signals should represent the actual scan scenarios including scan acceleration and system imperfections. As shown in our simulation and example in vivo results, the measurement errors acquired from an actual MRF scan with spiral sampling cannot be simply modeled as Gaussian white noise. First, as shown in Fig. 5, the aliasing artifacts from the in vivo scan have a structure-dependent spatial distribution; and the magnitude of the aliasing artifacts are much higher than the random thermal noise. Second, background phase that may be caused by system imperfections is also essential to include in the simulation to replicate the artifacts from in vivo images, whereas Gaussian noise is not able to represent the background phase. The interference of spatially dependent phase errors and aliasing artifacts makes the signal modeling difficult. Direct simulation of aliasing artifacts and background phases gives a more realistic approximation to signal distortions acquired in vivo.

The proposed method synthesizes aliased images from MRF scans acquired with high undersampling factors and system imperfections that closely resemble actual scan conditions. Under the assumption that a finite number of tissues are present in the brain, the aliased images can be calculated as a product of the spatial response functions from any undersampling patterns and the temporal functions from time-varying acquisition designs. This method allows fast simulation of undersampling artifacts in MR acquisition without compromising the accuracy of approximations. Comparison of signals and MRF maps simulated using the proposed method with those acquired from an in vivo scan justifies that our approach closely approximates the aliasing artifacts in signals and images acquired in actual accelerated MRF scans, even though a simplified phantom with limited numbers of tissue types is used. Upgrading the phantom with 7 more tissue components and partial volume effects does not cause significant difference in the simulations, since WM, GM and CSF occupy the majority of central tissue areas on a brain image. Adding a thin layer of extra tissues on the boundaries outside of the region of interest minimally alters the spatial response. In cases where disease conditions need to be modelled, the brain phantom could be modified to accommodate the changes in spatial distributions and tissue properties due to large pathological lesions.

The proposed simulator accelerates the calculations by a factor of 158 for an MRF scan simulation with 480 images, as compared to conventional methods involving gridding and NUFFT reconstruction. Such acceleration benefits from the spatiotemporal incoherence of MRF signals. The computationally expensive NUFFT calculations only involve spatial components that remain constant throughout different MRF scans, including anatomical tissue distributions, background field inhomogeneity, and sampling trajectories, thus the spatial responses can be extracted and pre-evaluated for repeated use. The example MRF framework in this study employed 48 linear ordered spiral trajectories and 3 segmented tissue distribution masks, leading to 144 spatial response functions in total. More variations in sampling patterns and more complex tissue compositions will require more spatial response functions to be precalculated. When a series of undersampled phantom images need to be synthesized based on an acquisition pulse sequence, the only computations left are the transient signal evolutions. Such strategy is especially useful in the context of MRF sequence optimization. Given so many pulse variables that constitute a very complex cost function, the optimization landscape is highly dimensional, non-convex and discontinuous. Searching for the optimal design using stochastic algorithms requires tens of thousands of iterations of cost function evaluations that involve image series simulations. The proposed fast simulator enables us to spare the excess efforts to iteratively compute NUFFT during optimization processes, otherwise accurately modelling actual image artifacts in the optimization is not realistic. And the optimization framework presented here that used the proposed simulator to minimize aliasing errors is proven to be capable of designing MRF scans of high measurement accuracy.

The benefits of the proposed fast simulation model lie in its great flexibility that can be applied to any sampling schemes, including Cartesian and any arbitrary trajectories. This model is also capable of integrating other spatially-dependent factors. In this study we demonstrated integrating the background field or receive coil phase. Other feasible components that can be simulated include random noise, or nonlinear ordering of spiral trajectories.

## VII. CONCLUSION

We propose a fast MRF image simulation model that provides accurate and accelerated simulation of undersampled MR images. The acceleration is more significant when a large amount of aliased images need to be simulated and evaluated from different sequence designs, anatomical structures, sampling patterns and system conditions. One application presented here is MRF sequence optimization, in which the estimation errors of hundreds of thousands of sequence candidates need to be evaluated. To generate aliased images from different sequence candidates, only the temporal functions of a finite number of tissues need to be updated, while the spatial response functions stay unchanged. Such an approach substantially accelerates the optimization by avoiding the time consuming image reconstruction step. Another application may involve studies that implement deep learning for MRF reconstruction [21]–[23], where a large quantity of data that represent actual scan scenarios are needed for training the neural network. The proposed model enables efficient simulation of images with artifacts that closely approximate the actual in vivo scans, which could potentially improve the robustness of the deep learning framework and reduce the need for actually acquired data.

## ACKNOWLEDGMENT

The authors would like to acknowledge funding from Siemens Healthineers, NIH grant EB026764-01 and NS 109439-01.

## REFERENCES

[1] D. Ma *et al.*, "Magnetic Resonance Fingerprinting," *Nature*, vol. 495, no. 7440, pp. 187–192, 2013.
[2] C. Liao *et al.*, "3D MR fingerprinting with accelerated stack-of-spirals and hybrid sliding-window and GRAPPA reconstruction," *Neuroimage*, vol. 162, pp. 13–22, Nov. 2017, doi: 10.1016/j.neuroimage.2017.08.030.
[3] D. Ma *et al.*, "Fast 3D magnetic resonance fingerprinting for a whole-brain coverage," *Magn. Reson. Med.*, vol. 79, no. 4, pp. 2190–2197, Apr. 2018, doi: 10.1002/mrm.26886.
[4] X. Cao, H. Ye, C. Liao, Q. Li, H. He, and J. Zhong, "Fast 3D brain MR fingerprinting based on multi-axis spiral projection trajectory," *Magn. Reson. Med.*, vol. 82, no. 1, pp. 289–301, Jul. 2019, doi: 10.1002/mrm.27726.
[5] B. Zhao *et al.*, "Optimal experiment design for magnetic resonance fingerprinting: Cramér-rao bound meets spin dynamics," *IEEE*

*Trans. Med. Imaging*, vol. 38, no. 3, pp. 844–861, Mar. 2019, doi: 10.1109/TMI.2018.2873704.

[6] O. Cohen and M. S. Rosen, "Algorithm comparison for schedule optimization in MR fingerprinting," *Magn. Reson. Imaging*, vol. 41, pp. 15–21, Sep. 2017, doi: 10.1016/j.mri.2017.02.010.

[7] K. Sommer, T. Amthor, M. Doneva, P. Koken, J. Meineke, and P. Börnert, "Towards predicting the encoding capability of MR fingerprinting sequences," *Magn. Reson. Imaging*, vol. 41, pp. 7–14, Sep. 2017, doi: 10.1016/j.mri.2017.06.015.

[8] C. C. Stolk and A. Sbrizzi, "Understanding the Combined Effect of k-Space Undersampling and Transient States Excitation in MR Fingerprinting Reconstructions," *IEEE Trans. Med. Imaging*, vol. 38, no. 10, pp. 2445–2455, Oct. 2019, doi: 10.1109/TMI.2019.2900585.

[9] J. A. Fessler and B. P. Sutton, "Nonuniform Fast Fourier Transforms Using Min-Max Interpolation," 2003.

[10] S. Kirkpatrick, C. D. Gelatt, and M. P. Vecchi, "Optimization by simulated annealing," *Science (80-. ).*, vol. 220, no. 4598, pp. 671–680, 1983, doi: 10.1126/science.220.4598.671.

[11] Z. P. Liang, "Spatiotemporal imaging with partially separable functions," in *2007 4th IEEE International Symposium on Biomedical Imaging: From Nano to Macro - Proceedings*, 2007, pp. 988–991, doi: 10.1109/ISBI.2007.357020.

[12] F. Lam and Z. P. Liang, "A subspace approach to high-resolution spectroscopic imaging," *Magn. Reson. Med.*, vol. 71, no. 4, pp. 1349–1357, 2014, doi: 10.1002/mrm.25168.

[13] A. G. Christodoulou *et al.*, "High-resolution cardiac MRI using partially separable functions and weighted spatial smoothness regularization," in *2010 Annual International Conference of the IEEE Engineering in Medicine and Biology Society, EMBC'10*, 2010, pp. 871–874, doi: 10.1109/IEMBS.2010.5627889.

[14] X. Feng *et al.*, "A robust algorithm for high-resolution dynamic MRI based on the partially separable functions model," *Magn. Reson. Imaging*, vol. 30, no. 5, pp. 620–626, Jun. 2012, doi: 10.1016/j.mri.2012.02.004.

[15] "BIC - The McConnell Brain Imaging Centre: ICBM 152 N Lin 2009." [Online]. Available: http://www.bic.mni.mcgill.ca/ServicesAtlases/ICBM152NLin2009. [Accessed: 08-Apr-2020].

[16] Y. Jiang, D. Ma, N. Seiberlich, V. Gulani, and M. A. Griswold, "MR fingerprinting using fast imaging with steady state precession (FISP) with spiral readout," *Magn. Reson. Med.*, vol. 74, no. 6, pp. 1621–1631, Dec. 2015, doi: 10.1002/mrm.25559.

[17] J. P. Wansapura, S. K. Holland, R. S. Dunn, and W. S. Ball, "NMR relaxation times in the human brain at 3.0 Tesla," *J. Magn. Reson. Imaging*, vol. 9, no. 4, pp. 531–538, Apr. 1999, doi: 10.1002/(SICI)1522-2586(199904)9:4<531::AID-JMRI4>3.0.CO;2-L.

[18] D. Kara, M. Fan, J. Hamilton, M. Griswold, N. Seiberlich, and R. Brown, "Parameter map error due to normal noise and aliasing artifacts in MR fingerprinting," *Magn. Reson. Med.*, vol. 81, no. 5, pp. 3108–3123, May 2019, doi: 10.1002/mrm.27638.

[19] M. Jarret and B. Lackey, "Substochastic Monte Carlo Algorithms," Apr. 2017.

[20] S. P. Jordan *et al.*, "Automatic Design of Sequence Pulses for Magnetic Resonance Fingerprinting using Physics-Inspired Optimization," 2021.

[21] O. Cohen, B. Zhu, and M. S. Rosen, "MR fingerprinting Deep RecOnstruction NEtwork (DRONE)," *Magn. Reson. Med.*, vol. 80, no. 3, pp. 885–894, Sep. 2018, doi: 10.1002/mrm.27198.

[22] E. Hoppe *et al.*, "Deep Learning for Magnetic Resonance Fingerprinting: A New Approach for Predicting Quantitative Parameter Values from Time Series," 2017, doi: 10.3233/978-1-61499-808-2-202.

[23] Z. Fang, Y. Chen, W. Lin, and D. Shen, "Quantification of relaxation times in MR Fingerprinting using deep learning.," *Proc. Int. Soc. Magn. Reson. Med. Sci. Meet. Exhib. Int. Soc. Magn. Reson. Med. Sci. Meet. Exhib.*, vol. 25, Apr. 2017.